# Evidence of defect phase formation in solid Xe induced by synchrotron radiation

*N.Yu. Masalitina, O.N. Bliznjuk, A.N. Ogurtsov*

*National Technical University "KhPI", Frunse Street 21, Kharkov 61002, Ukraine*

Rare-gas solids are the model systems in physics and chemistry of solids, and a lot of information about dynamics of electronic excitations, in particular, about subthreshold inelastic radiation-induced atomic processes, has been documented in several books and reviews (see e.g. Ref. 1 and references therein). It is generally recognized that creation of electronic excitations in rare-gas solids results in self-destruction of crystal lattice [1]. The investigation of the physics and chemistry of radiation-induced elementary processes of conversion of electronic excitations into structural defects in rare-gas solids is of considerable interest from the point of view of material modification by electronic excitations [2].

Contrary to conventional chemical or material engineering, which is mostly applied high temperature and/or high-pressures processes for material synthesis/modification and quite often the catalysts to speed up the reaction, the radiation is the unique source of energy, which can initiate chemical reactions at any temperature, under any pressure, in any phase, without use of catalysts. The subthreshold inelastic elementary processes in rare-gas solids induced by irradiation by synchrotron radiation such as defect formation and desorption under excitation by particles with kinetic energies below the threshold of knock-on of atoms from the lattice sites were studied recently [1]. Our study revealed that these processes have general similarity, they are all pass through the stage of trapping or self-trapping of mobile electronic excitations and, in principle, may be considered within the context of a single kinetic model [3], and that the harnessing of the intrinsic luminescence of rare-gas solids allows to carry out the real-time monitoring of the crystal structure of the samples [4]. Exciton self-trapping induced defect accumulation progress curves – time dependence of intensity of "defect" subband $M_1$ [1] – show pronounced linear growth at the beginning of irradiation, which follows by saturation at longer times. However, if irradiation of the sample keeps running, the decrease of the intensity of the "defect" subband was detected [5], which was much slower than initial growth.

In the present study we apply simple kinetic approach to such a progress curves and propose an explanation of the slow decay of the $M_1$-subband as a result of defect phase formation induced by Frenkel pairs creation during relaxation of self-trapped excitons in solid Xe. The experiments were carried out at the SUPERLUMI-station at HASYLAB, DESY, Hamburg. The selective photon excitation was performed with spectral resolution $\Delta\lambda = 0.2$ nm. The VUV-luminescence analysis was performed with $\Delta\lambda = 2$ nm, Pouey high-flux monochromator equipped with a multisphere plate detector.

Figure 1 shows example of dose dependence of intensity of $M_1$-subband of solid Xe under irradiation by photons with energies $E < E_g$. The luminescence spectra were measured under steady state irradiation conditions, using a single-photon counting mode with accumulation time $t = 1$ s in every point of measurement, which allows to exclude the influence of pulsed nature of synchrotron radiation on progress curves of defect accumulation. Initial increase of the intensity of the defect component during irradiation reflects the accumulation of stable long-lived point defects (Frenkel pairs) in the lattice as a result of exciton creation and self-trapping in the consecutive process $E + T \leftrightarrow MTE \rightarrow D$ [4], where $E$ is the mobile excitation (free exciton), which is trapped at trapping center $T$ (lattice imperfection) and forms an excited metastable trapped center $MTE$ (M-STE). Radiative decay of the $MTE$-center either returns the lattice into the initial state without permanent defect, or forms the permanent defect $D$ (Frenkel pair).

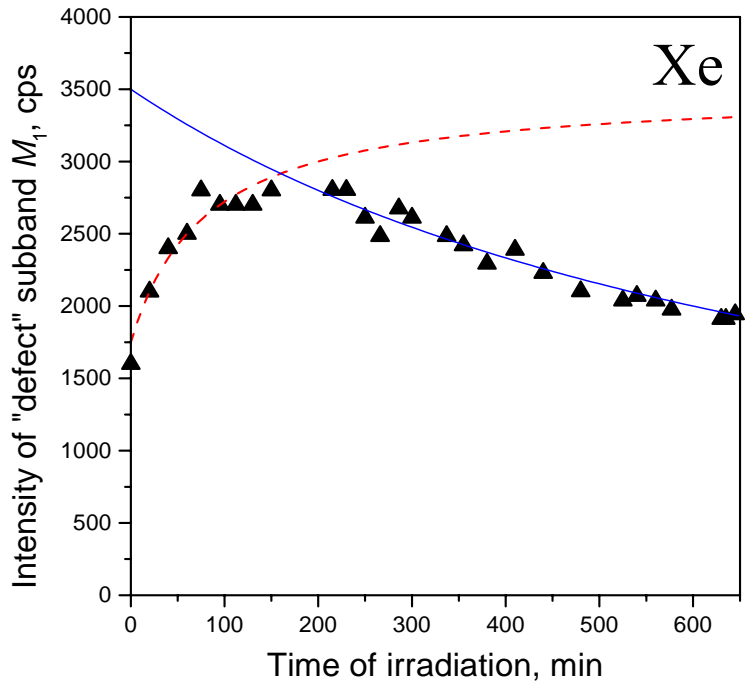

Figure 1: Dose dependence of intensity of $M_1$-subband of solid Xe under irradiation by photons with $h\nu = 9.15$ eV at $T = 20$ K and its fitting by Eq. 1 (dash curve) and Eq. 2 (solid curve).

The time dependence of luminescence intensity of $M_1$ subband under steady-state conditions may be expressed in form:

$$I(t) = I_0 + K \cdot t \cdot (L+t)^{-1}, \qquad (1)$$

where $I_0 = I(0)$ is the initial intensity of "defect" luminescence; $K$ is the saturation value of $(I(t) - I_0)$ at $t \to \infty$; $L \sim n_E n_T (n_{MTE})^{-1}$ is a characteristic constant of a sample – under identical excitation and detection conditions the sample with less pronounced processes of defect formation will have a bigger value of $L$. At high defect concentration, $n_D$, the exciton self-trapping near existing defects will not produce the separate point defect, but will induce the aggregation of defects and growth of the defect phase into the crystal in the process $MTE + D \to DD$. The time dependence in this case may be expressed in form:

$$I(t) = K \cdot L' \cdot (L' + t)^{-1}, \qquad (2)$$

where $L' \sim n_{MTE} n_D (n_{DD})^{-1}$. The best fit of our data results in values $K = 3500$ cps, $L = 40$ min, $L' = 800$ min. Analytical applications of this method enable an estimation of the "defectiveness" of the sample and maximum radiation dose, above which the growth of the defect phase occurs in the sample.

**References**

[1] A.N. Ogurtsov, *Advances in Spectroscopy of Subthreshold Inelastic Radiation-Induced Processes in Cryocrystals*, in: E.C. Faulques et al. (eds.), Spectroscopy of Emerging Materials, Kluwer Academic Publishers, Dordrecht (2004).
[2] N. Itoh, A.M. Stoneham, *Materials Modification by Electronic Excitation*, Cambridge University Press, Cambridge (2001).
[3] H. Schmalzried, *Chemical Kinetics of Solids*, VCH Verlag, Weinheim (1995).
[4] A.N. Ogurtsov, N.Yu. Masalitina, O.N. Bliznjuk, Low Temp. Phys. (2007) in press.
[5] A.N. Ogurtsov, E.V. Savchenko, M. Kirm, B. Steeg, G. Zimmerer, HASYLAB Jahresbericht 1997 (1998) 236.